\documentstyle[epsf,twocolumn,aps]{revtex}
\begin{document}
\title{On the Coexistence in RuSr$_2$GdCu$_2$O$_8$ 
of Superconductivity and Ferromagnetism}
\author{R. Weht, A. B. Shick, and W. E. Pickett}
\address{Department of Physics, University of California,
Davis CA 95616}

\maketitle

\begin{abstract}
We review the reasons that make superconductivity unlikely to arise
in a ferromagnet.  Then, in light of the report by Tallon and
collaborators that RuSr$_2$GdCu$_2$O$_8$ becomes superconducting at
$\sim$35 K which is well below the Curie temperature of 132 K, we 
consider whether the objections really apply to this compound.  Our
considerations are supported by local spin density calculations for
this compound, which indeed indicate a ferromagnetic RuO$_2$ layer.
The Ru moment resides in
$t_{2g}$ orbitals but is characteristic of itinerant magnetism (and
is sensitive to choice of exchange-correlation potential and to the
atomic positions).  Based on the small exchange splitting that is
induced in the Cu-O layers,
the system seems capable of supporting singlet superconductivity 
an FFLO-type order parameter and possibly a 
$\pi$-phase alternation between layers.  If instead the pairing is 
triplet in the RuO$_2$ layers, it can be distinguished by a
spin-polarized supercurrent.  Either type of superconductivity
seems to imply a spontaneous vortex phase if the magnetization is
rotated out of the plane.
\end{abstract}

\section*{Introduction}
Tallon and collaborators (\cite{tallon,bernhard} report the
remarkable observation of superconductivity arising up to at
least T$_s$ = 35 K in RuSr$_2$GdCu$_2$O$_{8-\delta}$ (Ru1212),
a compound reported originally by Bauerfeind {\it et al.},\cite{bauern}
in spite of
its having become ferromagnetic (FM) already at T$_m$ = 132 K.  The
ferromagnetism is evident from the magnetization vs. field curve and
also from zero-field muon spin rotation data that indicate a uniform
field in the sample below T$_m$.  Superconductivity
(SC) is evident from the loss of resistivity, reversal of the
susceptibility, and from specific heat measurements that reflect a
bulk transtion at T$_s$.  Although Felner {\it et al.} have reported
related results in $R_{1.4}$Ce$_{0.6}$Sr$_2$GdCu$_2$O$_{10}, R$ =
Eu or Gd,\cite{feln1}, 
they indicate that their system
is a canted antiferromagnet, with a saturation field an order of
magnitude smaller than Ru1212.  Thus Ru1212 appears to be unique as
a FM that becomes superconducting well within the FM phase.

Ru1212 presents scientific questions on several levels (not to mention 
the novel applications that a superconducting FM might have).  There
are questions on the phenomenological level: how can SC and FM coexist?
does the magnetic field arising from the frozen magnetization lead to
supercurrent flow? is the SC pairing singlet or triplet?  There are
also questions on the microscopic level: what is the FM like -- is it Ru, as
interpreted so far? how much does FM affect the carriers in the CuO$_2$
planes? how does $c$-axis dispersion compare with other cuprates?  In
this paper we begin to address both the phenomenological and the 
microscopic questions.

\section*{Questions of Coexistence of Superconductivity with Ferromagnetism}
The first question, and easiest to deal with, is the possibility of
paramagnetic limiting.  The observed saturation magnetization is about
1 $\mu_B$ per unit cell, which translates into an internal field $B_{int}
=4\pi M$ = 700 G.  Spin splitting 2$g\mu_B B_{int}$ of the two
electrons of the pair is negligible 
compared to the superconducting gap 2$\Delta \sim 4k_BT_c$, so paramagnetic
limiting is no problem.

Since there is a strong tendency for singlet pairing in materials
with CuO$_2$ layers such as Ru1212 has, and substitution of Zn for Cu
strongly reduces T$_s$ as in cuprates,
we examine the possibility of a singlet SC (sSC) state in FM Ru1212
by considering two fundamental problems: (1) 
how does the SC order parameter accommodate itself to the vector
potential arising from the intrinsic magnetization, and
(2) what type of pairing occurs when the exchange field splits majority and
minority Fermi surfaces?   We
find that it is primarily the latter item  (exchange splitting)
that presents difficulty for singlet SC in this system,
and we consider briefly whether the problem can be alleviated by
development of a
FFLO phase\cite{fflo} or by ``$\pi$ phase'' formation.\cite{prokic}
We note further that triplet pairing in the RuO$_2$ layer, which appears to
occur in Sr$_2$RuO$_4$, should not be forgotten.  If Ru1212 is instead
a triplet superconductor (in the Ru plane), it would have several
exciting characteristics, such as a polarized supercurrent.

\section*{Issues of Coexistence}
Early on, Ginzburg\cite{ginzburg} observed that, although constant bulk
magnetization $M$ does not induce supercurrent flow, the vanishing of 
$M$ at the boundaries induces surface currents that try to shield
the exterior from the internal magnetic field $B_{int}=4\pi M$ (inverse
Meissner effect).  This
increase in energy would suppress the SC state unless the sample cross
secion is not much larger than the penetration depth.  In type II
superconductors, however, it was observed by Krey\cite{krey} that a
spontaneous vortex phase (SVP) can be formed that avoids Ginzburg's
difficulty.  Since that time there have been studies of the competition
between FM and SC,\cite{FMSC} including suggestions that
a SVP may have been observed,
but almost always for the case where T$_m$ is
less than but comparable to T$_s$.

One exception to this regime is the
recent work of Felner and collaborators,\cite{feln1,feln2} who suggest 
that a SVP may occur in the canted antiferromagnet
$R_{1.4}$Ce$_{0.6}$Sr$_2$GdCu$_2$O$_{10}, R$ =
Eu and Gd.  Ru1212 is different in two ways: (1)
the saturation field is itself a factor of ten larger; however, this is
of limited importance because Pauli limiting is not a factor in either
type of material, and (2) Ru1212, being FM, has an exchange splitting
of the order of 1 eV
(we show below) that will induce an exchange splitting of the
paired electrons in the CuO$_2$ planes and thereby cause additional
difficulty for SC pairing.

Tallon {\it et al.} expect the magnetization $M$ to lie in
the plane, based on the magnitude of the field at the muon site observed
in zero field $\mu$SR studies.\cite{tallon} If this is the case, it
reduces some of the coexistence questions, because orbital kinetic energy
change of the electrons (or pairs) would require the (very small) $c$
axis hopping, and thus is expected to be negligible compared
to the gap energy.  Then the remaining effect of the internal field is
to lift the degeneracy of the up and down spin electrons (Zeeman 
splitting).  This effect makes the up and down Fermi surfaces 
inequivalent and causes singlet pairs to have a net momentum 
$Q\equiv k_{F\uparrow}-k_{F\downarrow}\ne$0.
The result can be a state of the type first discussed by Fulde and
Farrel and by Larkin and Ovchinnikov\cite{fflo} (FFLO), in which the
SC order parameter becomes inhomogeneous to accommodate the non-zero
momentum pairs.  Another way to decrease the competition between SC and
FM is to form a $\pi$ phase in which the SC order parameter vanishes in
the FM Ru layer.\cite{prokic}

Burkhardt and
Rainer\cite{rainer} have made an extensive study of the two 
dimensional superconductor in a magnetic field, and they find that an
FFLO phase is preferred over a homogeneous SC state above a lower 
critical field $B_{c1}$, but only well below T$_s$ and at relatively
high field.  To some extent their analysis may be applicable
to Ru1212, with the change that the
``magnetic field'' is frozen in and arises from electronic exchange.
The exchange splitting, even if small on an electronic energy 
scale, may correspond 
to a very large field, and its magnitude is one feature we address below.

\section*{Microscopic Electronic and Magnetic Structure}
\subsection*{Electronic Structure Methods}
To address the question of magnetism in Ru1212
we have applied both the local density approximation (LDA)\cite{cepald}
and the `semilocal' generalized gradient approximation (GGA)\cite{gga}
to describe the effects of exchange and correlation.  The predictions
actually differ considerably, and we report only the GGA results here.
Calculations were done using the linearized augmented plane wave (LAPW) 
method\cite{djsbook,wien,details} 
to find the electronic and magnetic ground state, the band structure
and projected densities of states (PDOS). 

\subsection*{Crystal Structure}
The structure of Ru1212, of the triple-perovskite type,
is comprised of double
CuO$_2$ (O$_{Cu}$ site) layers separated by a Gd layer, sandwiched in
turn by SrO (O$_{apical}$ site) layers.
The unit cell in completed by a RuO$_2$ (O$_{Ru}$ site) layer,
making it structurally
related to YBa$_2$Cu$_3$O$_7$.  The difference is that the CuO chain layer is
replaced by a RuO$_2$ square planar layer, with resulting tetragonal
symmetry.  The oxygen variability that occurs lies primarily in the
Sr layers, and there are displacements from the ideal positions\cite{tallon}
for all O ions, with the displacements being largest for O$_{Ru}$.

\begin{figure}[b!] 
\epsfxsize=4.5cm\centerline{\epsffile{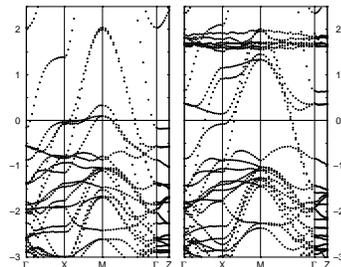}}
\caption{Majority (left) and minority (right) bands near E$_F$ in tetragonal
RuSr$_2$GdCu$_2$O$_8$ along high symmetry directions.  Note that the
distinctive Cu-O bands that reach their maximum at M lie at the same
position for both spins, whereas the Ru bands are split by $\sim$1 eV.
The flat bands near 2 eV are the minority Gd 4$f$ states.
}
\label{f1:fig1}
\end{figure}

\subsection*{Discussion of Results}
Gd behaves as a magnetic trivalent ion as expected, with moment very
close to 7 $\mu_B$.  In expectation that the Gd moment has little
effect on the electronic and magnetic behavior in the rest of the cell,
we have treated only a FM alignment of Gd ions.  (They actually order
antiferromagnetically around 2.6 K, but antiferromagnetic order would require
doubling of the unit cell with no gain of information or insight.) 
We have found that 
the electronic and magnetic structure in the cuprate and ruthenate layers does
not depend on
whether the Gd moment is parallel or antiparallel to the Ru moment.

We obtain a FM Ru layer, with the band structure shown in Fig. 1 and the
relevant projected densities of states presented in Fig. 2.  The moment
(besides that on Gd) is $\sim2.5 \mu_B$ per cell, about 1.5 $\mu_B$ of 
which is directly assignable to the Ru ion.  The remainder is spread
among the six neighboring O ions, all of which are strongly polarized 
(for a nominally O$^{2-}$ ion).  Our total moment is substantially more
than the 1 $\mu_B$ reported by Tallon {\it et al}.  
As mentioned above, they have 
evidence of displacements of the O ions in the Ru layer from their ideal 
positions by as much as 0.4 \AA.  We have observed in our calculations 
strong sensitivity of the moment to the position of the apical O and to the
distance between the Ru and Cu atoms, so it is quite possible that a more
realistic structure will give a moment much closer to the observed value.
Stoichiometry will be important as well.

\begin{figure}[b!] 
\epsfxsize=4.5cm\centerline{\epsffile{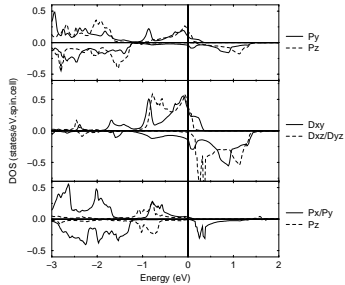}}
\vspace{10pt}
\caption{Atom projected densities of states for O$_{Ru}$ (top), Ru
(middle), and O$_{apical}$ (bottom), clearly indicating spin splitting
in all Ru $t_{2g}$ states and all $p$ orbitals of O$_{Ru}$, but in
only the $p_x, p_y$ orbitals of O$_{apical}$.
}
\label{f2:fig2}
\end{figure}

The Ru moment arises within the Ru $t_{2g}$ subshell.  The charge state
is difficult to specify due to the metallic character of the Ru-O layer, 
but it is {\it not}
close to Ru$^{5+}$ and may be closer to Ru$^{4+}$.  The exchange splitting
of the Ru $t_{2g}$ bands is 1 eV, but the induced splitting in the CuO$_2$
layers is close to two orders of magnitude smaller.  The very small
coupling is due to the fact that the Ru $t_{2g}$ states couple only to the
O$_{apical}~p_x, p_y$ orbitals, and these to not couple either to the
Cu $d_{x^2-y^2}$ or $s$ orbitals, so the coupling path is more indirect
(perhaps O$_{Ru}$-O$_{apical}$-O$_{Cu}$).  The induced moment is roughly
0.01 $\mu_B$ per CuO$_2$ layer, which leads to a displacement of spin up
and spin down barrel Fermi surfaces of only $\delta k_F \approx 0.01 k_F$.
The work of Burkhardt and Rainer\cite{rainer} suggest that a model independent
feature of FFLO-like solutions is that the wavelength of modulation of
the superconducting order parameter $\lambda_Q = 2\pi/\delta k_F$ must
be greater than (not necessarily comparable to) the coherence length
$\xi$.  Our estimate is $\lambda_Q \sim$ 140$a$, which is much greater
than representative cuprate values $\xi \sim 5-15a$.  Hence the
exchange splitting we obtain is small enough to allow the
possibility of an FFLO-like phase in Ru1212.

\section*{Summary and Acknowledgments}
With this paper we have begun the study of the superconducting ferromagnet
Ru1212.  Although magnetism in the Ru layer is strong, the exchange
coupling to the Cu layers is quite small.  The most likely scenario seems
to be an FFLO-like inhomogeneous SC order parameter in the cuprate layers.

We acknowledge important close communication with J. Tallon and a
preprint of~\cite{bernhard} from C. Bernhard.
This research was supported by Office of Naval Research
Grant No. N00014-97-1-0956 and National Science foundation Grant
DMR-9802076.  Computation was supported by the National Partnership for
Advanced Computational Infrastructure (NPACI).


\begin{references}
\bibitem{details}The sphere radii
used in fixing the LAPW basis were chosen to be 2.00 a.u for Sr and Gd,
1.9 for Ru and Cu,
and 1.60 for N. Local orbitals for semicore states and for Gd $4f$
were added to the basis set for extra
flexibility and to allow semicore states to be treated within the same
energy window as the band states.  The LAPW plane wave
cutoff corresponded to energy of 14 Ryd, making the matrix dimension
approximately 1400.
\end{references}
\end{document}